\documentclass[twocolumn]{aastex63}

\usepackage{lineno}

\submitjournal{ApJ}

\shorttitle{Surface Brightness Profile of \Lya Halos out to $320$~kpc in HETDEX}
\shortauthors{Lujan Niemeyer et al.}
\graphicspath{{./}{figures/}}

\usepackage{amsmath}
\newcommand{\angstrom}{\textup{\AA}}
\newcommand{\Lya}{\mbox{Lyman-$\alpha$} }

\begin{document}

\title{Surface Brightness Profile of \Lya Halos out to $320$~kpc in HETDEX}

\author[0000-0002-6907-8370]{Maja Lujan Niemeyer}
\affiliation{Max-Planck-Institut f\"{u}r Astrophysik, Karl-Schwarzschild-Str. 1, 85741 Garching, Germany}
\email{maja@mpa-garching.mpg.de}

\author[0000-0002-0136-2404]{Eiichiro Komatsu}
\affiliation{Max-Planck-Institut f\"{u}r Astrophysik, Karl-Schwarzschild-Str. 1, 85741 Garching, Germany}
\affiliation{Kavli Institute for the Physics and Mathematics of the Universe (Kavli IPMU, WPI), University of Tokyo, Chiba 277-8582, Japan}

\author[0000-0002-0885-8090]{Chris Byrohl}
\affiliation{Max-Planck-Institut f\"{u}r Astrophysik, Karl-Schwarzschild-Str. 1, 85741 Garching, Germany}

\author[0000-0002-8925-9769]{Dustin Davis}
\affiliation{Department of Astronomy, The University of Texas at Austin, 2515 Speedway Boulevard, Austin, TX 78712, USA}

\author[0000-0002-7025-6058]{Maximilian Fabricius}
\affiliation{Max-Planck-Insitut f\"ur extraterrestrische Physik, Giessenbachstrasse, 85748 Garching, Germany}
\affiliation{Universit\"ats-Sternwarte M\"unchen, Scheinerstrasse 1, D-81679 München, Germany}

\author[0000-0002-8433-8185]{Karl Gebhardt}
\affiliation{Department of Astronomy, The University of Texas at Austin, 2515 Speedway Boulevard, Austin, TX 78712, USA}

\author[0000-0001-6717-7685]{Gary J. Hill}
\affiliation{McDonald Observatory, The University of Texas at Austin, 2515 Speedway Boulevard, Austin, TX 78712, USA}
\affiliation{Department of Astronomy, The University of Texas at Austin, 2515 Speedway Boulevard, Austin, TX 78712, USA}

\author[0000-0003-2977-423X]{Lutz Wisotzki}
\affiliation{Leibniz-Institut f\"ur Astrophysik Potsdam (AIP), An der Sternwarte 16, D-14482 Potsdam, Germany}

\author[0000-0003-4381-5245]{William P. Bowman}
\affiliation{Department of Astronomy \& Astrophysics, The Pennsylvania State University, University Park, PA 16802, USA}
\affiliation{Institute for Gravitation and the Cosmos, The Pennsylvania State University, University Park, PA 16802, USA}

\author[0000-0002-1328-0211]{Robin Ciardullo}
\affil{Department of Astronomy \& Astrophysics, The Pennsylvania
State University, University Park, PA 16802, USA}
\affil{Institute for Gravitation and the Cosmos, The Pennsylvania
State University, University Park, PA 16802, USA}

\author[0000-0003-2575-0652]{Daniel J. Farrow}
\affiliation{Max-Planck-Insitut f\"ur extraterrestrische Physik, Giessenbachstrasse, 85748 Garching, Germany}
\affiliation{Universit\"ats-Sternwarte M\"unchen, Scheinerstrasse 1, D-81679 München, Germany}

\author[0000-0001-8519-1130]{Steven L. Finkelstein}
\affiliation{Department of Astronomy, The University of Texas at Austin, 2515 Speedway Boulevard, Austin, TX 78712, USA}

\author[0000-0003-1530-8713]{Eric Gawiser}
\affiliation{Rutgers, The State University of New Jersey, Piscataway, NJ 08854, USA}

\author[0000-0001-6842-2371]{Caryl Gronwall}
\affiliation{Department of Astronomy \& Astrophysics, The Pennsylvania State University, University Park, PA 16802, USA}
\affiliation{Institute for Gravitation and the Cosmos, The Pennsylvania State University, University Park, PA 16802, USA}

\author[0000-0002-8434-979X]{Donghui Jeong}
\affiliation{Department of Astronomy \& Astrophysics, The Pennsylvania State University, University Park, PA 16802, USA}
\affiliation{Institute for Gravitation and the Cosmos, The Pennsylvania State University, University Park, PA 16802, USA}

\author[0000-0003-1838-8528]{Martin Landriau}
\affiliation{Lawrence Berkeley National Laboratory, 1 Cyclotron Road, Berkeley, CA 94720, USA}

\author[0000-0001-5561-2010]{Chenxu Liu}
\affiliation{Department of Astronomy, The University of Texas at Austin, 2515 Speedway Boulevard, Austin, TX 78712, USA}

\author[0000-0002-2307-0146]{Erin Mentuch Cooper}
\affiliation{Department of Astronomy, The University of Texas at Austin, 2515 Speedway Boulevard, Austin, TX 78712, USA}
\affiliation{McDonald Observatory, The University of Texas at Austin, 2515 Speedway Boulevard, Austin, TX 78712, USA}

\author[0000-0002-1049-6658]{Masami Ouchi}
\affiliation{National Astronomical Observatory of Japan, 2-21-1 Osawa, Mitaka, Tokyo 181-8588, Japan}
\affiliation{Institute for Cosmic Ray Research, The University of Tokyo, 5-1-5 Kashiwanoha, Kashiwa, Chiba 277-8582, Japan}
\affiliation{Kavli Institute for the Physics and Mathematics of the Universe (Kavli IPMU, WPI), University of Tokyo, Chiba 277-8582, Japan}

\author[0000-0001-7240-7449]{Donald P. Schneider}
\affiliation{Department of Astronomy \& Astrophysics, The Pennsylvania State University, University Park, PA 16802, USA}
\affiliation{Institute for Gravitation and the Cosmos, The Pennsylvania State University, University Park, PA 16802, USA}

\author[0000-0003-2307-0629]{Gregory R. Zeimann}
\affiliation{Hobby-Eberly Telescope, University of Texas, Austin, Austin, TX, 78712, USA}

\begin{abstract}

We present the median-stacked \Lya surface brightness profile of 968 spectroscopically selected \Lya emitting galaxies (LAEs) at redshifts $1.9<z<3.5$ in the early data of the Hobby-Eberly Telescope Dark Energy Experiment (HETDEX). The selected LAEs are high-confidence \Lya detections with large signal-to-noise ratios observed with good seeing conditions (point-spread-function full-width-at-half-maximum $<1.4''$), excluding active galactic nuclei (AGN). The \Lya luminosities of the LAEs are $10^{42.4}-10^{43}\,\mathrm{erg}\,\mathrm{s}^{-1}$. We detect faint emission in the median-stacked radial profiles at the level of $(3.6\pm 1.3)\times 10^{-20}\,\mathrm{erg}\,\mathrm{s}^{-1}\,\mathrm{cm}^{-2}\,\mathrm{arcsec}^{-2}$ from the surrounding \Lya halos out to $r\simeq 160$ kpc (physical).
The shape of the median-stacked radial profile is consistent at $r<80\,\mathrm{kpc}$ with that of much fainter LAEs at $3<z<4$ observed with the Multi Unit Spectroscopic Explorer (MUSE), indicating that the median-stacked Lyman-$\alpha$ profiles have similar shapes at redshifts $2<z<4$ and across a factor of $10$ in \Lya luminosity. While we agree with the results from the MUSE sample at $r<80\,\mathrm{kpc}$, we extend the profile over a factor of two in radius. At $r>80\,\mathrm{kpc}$, our profile is flatter than the MUSE model.
The measured profile agrees at most radii with that of galaxies in the Byrohl et al. (2021) cosmological radiative transfer simulation at $z=3$. This suggests that the surface brightness of a Lyman-$\alpha$ halo at $r\lesssim 100$ kpc is dominated by resonant scattering of \Lya photons from star-forming regions in the central galaxy, whereas at $r > 100$ kpc it is dominated by photons from galaxies in surrounding dark matter halos.
\end{abstract}


\section{Introduction}
\label{sec:intro}

Large \Lya emission regions with sizes of tens to hundreds of kpc were first  found around high-redshift radio galaxies \citep[][]{mccarthy/etal:1987,vanojik/etal:1997,overzier/etal:2001,reuland/etal:2003}. Later, they were observed around
quasars \citep[][]{weidinger/etal:2004,weidinger/etal:2005,christensen/etal:2006,smith/etal:2009,goto/etal:2009,cantalupo/etal:2014,martin/etal:2014a,martin/etal:2014b,hennawi/etal:2015,arrigonibattaia/etal:2016,borisova/etal:2016,cai/etal:2018,arrigonibattaia/etal:2019,kikuta/etal:2019,zhang/etal:2020},
star-forming galaxies \citep[][]{smith/etal:2007,smith/etal:2008,shibuya/etal:2018},
and in galaxy-overdense regions \citep[][]{steidel/etal:2000,matsuda/etal:2004,matsuda/etal:2011,yang/etal:2010,cai/etal:2017}.
Measurements of the cross-correlation of \Lya intensity with quasar \citep[][]{croft/etal:2016,croft/etal:2018} or LAE positions \citep[][]{kakuma/etal:2021,kikuchihara/etal:2021} revealed \Lya emission on even larger scales.
Most individual high-redshift star-forming galaxies such as \Lya emitting galaxies (LAEs) and Lyman-break galaxies (LBGs) are surrounded by smaller, $1-10$ kpc-size \Lya halos \citep[][]{hayashino/etal:2004,swinbank/etal:2007,rauch/etal:2008,ouchi/etal:2009,wisotzki/etal:2016,patricio/etal:2016,smit/etal:2017,leclercq/etal:2017,erb/steidel/chen:2018,claeyssens/etal:2019,claeyssens/etal:2022}. 
\citet{kusakabe:etal/2022} detected \Lya halos with sizes between $10$ and $50\,\mathrm{kpc}$ in 17 out of 21 continuum-selected galaxies.
While \citet[][]{bond/etal:2010}, \citet[][]{feldmeier/etal:2013}, and \citet{jiang/etal:2013} report a lack of evidence for spatially extended \Lya emission around LAEs and LBGs, the ubiquitous presence of extended \Lya halos around high-redshift star-forming galaxies has been confirmed by stacking analyses \citep[][]{moeller/warren/1998,steidel/etal:2011,matsuda/etal:2012,momose/etal:2014,momose/etal:2016,xue/etal:2017,wisotzki/etal:2018,wu/etal:2020,huang/etal:2021}.
The Multi Unit Spectroscopic Explorer (MUSE, \citealt{bacon/etal:2010}) on the Very Large Telescope (VLT) has revolutionized the subject by using up to $31$ hour exposures to explore \Lya halos. By co-adding a sample of $270$ LAEs at $3 < z < 6$ in a $2\,\mathrm{arcmin}^2$ section of sky, \citet[][]{wisotzki/etal:2018} found that \Lya emission can be traced out to several arcseconds from the source centers, so that nearly all the sky is covered by \Lya emission around high-redshift galaxies in projection.

The size of the \Lya halos around LAEs depends on their physical properties such as the ultraviolet (UV) and \Lya luminosities and the size of the UV-emitting region \citep[][]{wisotzki/etal:2016,xue/etal:2017,momose/etal:2016,leclercq/etal:2017}. While \Lya halos are more compact in nearby galaxies than in high-redshift galaxies \citep[e.g.][]{hayes/etal:2005,oestlin/etal:2009,hayes/etal:2013,hayes/etal:2014,leclercq/etal:2017,rasekh/etal:2021}, no significant redshift evolution of halos around LAEs within $2<z<6$ has been detected \citep[][]{momose/etal:2014,leclercq/etal:2017,kikuchihara/etal:2021}. In contrast, observed \Lya halo profiles of quasars are shown to increase from $z\sim2$ to $z\sim3$ and remain constant from $z\sim3$ to $z\sim6$ \citep[][]{arrigonibattaia/etal:2016,arrigonibattaia/etal:2019,farina/etal:2019,cai/etal:2019,osullivan/etal:2020,fossati/etal:2021}.

There are various sources of \Lya photons that may contribute to the extended \Lya emission.
One substantial source of \Lya photons is the local recombination of hydrogen atoms ionized by photons from young, massive stars in star-forming galaxies or from active galactic nuclei \citep[AGN;][]{dijkstra:2019}.
Due to their resonant nature, \Lya photons are scattered by neutral hydrogen atoms in the circumgalactic medium (CGM) and the intergalactic medium (IGM).
\Lya photons can also be created by collisional excitation, such as when dense gas flows into a galaxy \citep[called ``gravitational cooling''][]{haiman/spaans/quataert:2000,fardal/etal:2001,fauchergiguere/etal:2010}, and by fluorescence  of hydrogen gas ionized by photons from more distant AGN or star-forming regions \citep[called the ``UV background'';][]{gould/weinberg:1996,cantalupo/etal:2005,kollmeier/etal:2010,mas-ribas/dijkstra:2016}. Satellite galaxies can also contribute to the extended \Lya emission \citep[][]{mas-ribas/etal:2017}.

In order to constrain the contributions of \Lya emission sources and mechanisms, it is necessary to model \Lya emission and its radiative transfer realistically.
One method to model radiative transfer in \Lya halos is a perturbative approach \citep[e.g.][]{kakiichi/etal:2018}. Another way uses hydrodynamical simulations that resolve the gas around galaxies, which can be post-processed with a Monte-Carlo radiative transfer calculation to predict the shape of \Lya emission around galaxies \citep[e.g.][]{lake/etal:2015,mitchell/etal:2021,kimock/etal:2021}.
Most of these models simulate small numbers of galaxies, while \citet{zheng/etal:2011}, \citet{gronke/bird:2017} and \citet{byrohl/etal:2021} calculate \Lya radiative transfer in cosmological hydrodynamical simulations and offer predictions for large samples of galaxies. While being a promising tool, hydrodynamical simulations of galaxy formation in cosmological volumes with sub-kpc resolutions \citep[e.g.][]{nelson/etal:2020} possibly suffer from convergence issues both in the physical gas state \citep[e.g.][]{vandevoort/etal:2019} and \Lya radiative transfer \citep[e.g.][]{camps/etal:2021}.
 
Comparisons of predictions with measurements draw different conclusions.
While \citet{steidel/etal:2011}, \citet{gronke/bird:2017}, and \citet{byrohl/etal:2021} find that most of the extended \Lya emission can be explained by scattering of \Lya photons from the central galaxy or nearby galaxies, \citet{lake/etal:2015} stressed the importance of cooling radiation in producing \Lya halos when they compared their simulation with observations from \citet{momose/etal:2014}. \citet{mitchell/etal:2021} report that satellite galaxies are the predominant source of \Lya photons at $10-40\,\mathrm{kpc}$, while cooling radiation also plays a relevant role.

Since the dominant origin of the \Lya halo photons depends on, among other things, the distance to the galaxies \citep[][]{mitchell/etal:2021,byrohl/etal:2021}, observations of \Lya profiles out to larger distances will be helpful.
An ideal data set for this is provided by the Hobby-Eberly Telescope Dark Energy Experiment \citep[HETDEX;][]{hill/etal:2008,gebhardt/etal:2021,hill/etal:2021}, which is designed to detect more than one million LAEs at redshifts $1.9<z<3.5$ in a $10.9 \,\mathrm{Gpc}^3$ volume to measure their clustering and thereby constrain cosmological parameters.
HETDEX detects emission lines by simultaneously acquiring tens of thousands of spectra without any pre-selecting of targets. 
In this work, we measure the median radial \Lya surface brightness profile of 968 LAEs at $1.9<z<3.5$ using the HETDEX data. We take advantage of the wide field of view of HETDEX and expand the measurement out to $320\,\mathrm{kpc}$ from the LAE centers.

This paper is structured as follows.
Section \ref{sec:data} describes the data, the data processing, and the definition of the LAE sample.
Section \ref{sec:methods} presents the method to obtain the median radial \Lya surface brightness profile of the LAEs and how we account for systematic errors.
Section \ref{sec:results} reports the results and shows that simple stacking of the \Lya surface brightness profiles reproduces the rescaled best-fit model of stacked \Lya halos at higher redshift ($3<z<4$) at $r<80\,\mathrm{kpc}$ from \citet{wisotzki/etal:2018}. We quantify the effect of possible AGN contamination in the LAE sample by stacking sources with broad lines and high luminosities.
In Section \ref{sec:discussion}, we compare our results with theoretical predictions from a radiative transfer simulation by \citet{byrohl/etal:2021} and from a perturbative approach by \citet{kakiichi/etal:2018}.
We conclude in Section \ref{sec:summary}.

We assume a flat $\Lambda$ cold dark matter ($\Lambda$CDM) cosmology consistent with the latest results from the {\it Planck} mission: $H_0=67.37\,\mathrm{km}\,\mathrm{s}^{-1}\,\mathrm{Mpc}^{-1}$ and $\Omega_{\mathrm{m},0}=0.3147$ \citep{planck/etal:2018}. All distances are in units of physical kpc/Mpc unless noted otherwise.

\section{Data and Galaxy Sample}\label{sec:data}
\subsection{Data and Data Processing}
\label{subsec:data_processing}
We use spectra from the internal data release 2.1.3 (DR 2.1.3) of HETDEX, which were obtained with the Visible Integral-field Replicable Unit Spectrograph (VIRUS) on the $10\,\mathrm{m}$ Hobby-Eberly Telescope (HET, \citealt{ramsey/etal:1994,hill/etal:2021}). VIRUS consists of up to 78 integral-field unit fiber arrays (IFUs), each of which contains 448 $1.5''$-diameter fibers and spans $51''\times 51''$ on the sky. The fibers from each IFU are fed to a low-resolution ($R=800$) spectrograph unit containing two spectral channels, which covers the wavelengths between $3500\,\angstrom$ and $5500\,\angstrom$. Each spectral channel has a CCD detector with two amplifiers; the spectra from the 448 fibers of each IFU are effectively split over four amplifiers. The IFUs with $\sim$ 35k total fibers are distributed throughout the $18'$ diameter of the telescope's field of view. Each HETDEX observation includes three 6-minute exposures, which are dithered to fill in gaps between the fibers. The IFUs are arrayed on a grid with $100''$ spacing. The gaps between the IFUs remain, so that the filling factor of one observation is $\sim 1/4.6$. Details of the upgraded HET and the VIRUS instrument can be found in \citet{hill/etal:2021}.

The data processing pipeline is described in \citet{gebhardt/etal:2021}.
A crucial aspect for detecting low surface brightness \Lya emission is the sky subtraction. HETDEX applies two separate approaches to sky subtraction, one using a local sky determined from a single amplifier (112 fibers spanning $\sim0.2\,\mathrm{arcmin}^2$), and a full-frame sky determined from the full array of IFUs (over 30k fibers).

For the local sky subtraction, we identify continuum sources by extracting a continuum estimate for each of the 112 fibers on a given amplifier using a wavelength range of $4100-5100\,\angstrom$. We flag fibers with a  $>3\sigma$ detection of continuum \citep[using a biweight scale as $\sigma$;][]{beers/etal:1990} as continuum fibers. We further flag the two adjacent fibers in the spectroscopic image to each continuum fiber. This typically removes about $15\%$ of the fibers. Of the remaining fibers, we apply a further cut of $10\%$ of the fibers with the highest counts in the continuum region. The biweight location, which is a robust estimate of the central location of a distribution \citep[][]{beers/etal:1990}, of the remaining approximately $75\%$ of the fibers determines the local sky spectrum. There is residual low-level background that is due to a combination of dark current, scattered light, mismatch of the fiber profile, and illumination differences for the specific exposure. After removing the identified continuum sources, we smooth the spectroscopic image with a two-dimensional biweight filter that is six fibers by $700\,\angstrom$ across in order to estimate and remove the broad-scale residual background (``background light correction''). 
This procedure is highly effective at removing the residual background at the expense of removing some continuum of faint sources. While this local sky subtraction is robust, extended emission that covers a significant fraction of the small area of an amplifier can be mistaken as sky emission and thus removed from the signal.

As an alternative, HETDEX provides a full-frame sky subtraction. This procedure uses over 30k fibers, which provides significant improvement for objects that dominate an amplifier.
Continuum sources are identified in the same manner as the local sky estimate.
The disadvantage of this procedure is that the amplifiers have their individual differences which need to be addressed.
These include differences in the illumination of the primary mirror on the IFUs across the $20'$ field, the wavelength solution, and the instrumental dispersion. 

There are two components to the amplifier-to-amplifier normalization for the full-frame sky spectrum. The first component is the instrumental throughput difference that we measure from the twilight frames. The second is due to the different illumination patterns of the primary as the HET tracks a specific field, leading to illumination differences of the IFUs. We use the relative residual flux in the sky-subtracted images to determine the relative normalizations due to illumination. Because there a broad wavelength dependence, we use a low-order term to adjust the normalization for each amplifier coming from $4$ wavelength regions averaged over $500\,\angstrom$ regions. The scalings range from $0.9$ to $1.1$, and we use deviations beyond this range as a flag for data that potentially has to be removed due detector controller issues. The wavelength dependence is small, generally under $1\%$. Since the local sky subtraction uses one sky spectrum for each amplifier, the illumination differences are irrelevant.

The other important aspect is due to small changes in the wavelength solution. We use the full-frame sky wavelength solution to adjust the local values. During the full-frame sky subtraction, we allow each amplifier to fit for a wavelength offset and a dispersion term. The offsets are generally small (less than $0.2\,\angstrom$), but the dispersion term can be more important. We think that the dispersion term is caused by differential breathing modes in the spectrograph due to temperature changes. For the majority of amplifiers, the sky-subtracted frames from the full frame and local sky look very similar. There are a few where the dispersion term does not adequately capture the changes in the wavelength solution, causing small residuals correlated with bright sky lines.
The background residual counts within a given amplifier still need to be removed, which will subtract some light from the faintest sources. We are confident that, while there remains some uncertainty in the background estimate, the line flux relative to the continuum is unaffected.

One aspect we do not account for are changes in instrumental resolution across spectrographs. While we have built these to have as uniform resolving power as possible, there are differences. In particular, the resolution can change across the amplifier, where fibers at the edge of spectrographs have larger instrumental dispersion. The local sky subtraction naturally deals with the variation from spectrograph to spectrograph, whereas the full-frame does not. The variation within a spectrograph is equally a problem in both sky subtraction procedures. We do not address these differences at this point. The effect is small, and only noticeable at some of the edges of spectrographs. To compensate for these issues, our noise model takes into account the larger residuals and increases the noise at those locations. The number of fibers affected by this increased noise is under 5\%.

The full-frame sky subtraction over the large field of view of VIRUS offers an advantage for extended objects over the local sky subtraction and over instruments such as MUSE, which has a smaller field of view with which to measure the sky spectrum.
We therefore adopt the full-frame sky subtraction.

\subsection{LAE Sample}\label{subsec:lae_sample}
We draw our sample of LAEs from the line emission catalog from the HETDEX internal data release 2 (specifically v2.1.3) to be published in Mentuch Cooper et al., in preparation, which contains approximately 300k LAEs.
For each detected line, the catalog provides the coordinates of the centroid, the central wavelength, the line width, the line flux, the signal-to-noise ratio ($S/N$), and the probability of the feature being a \Lya line. This probability is computed by the HETDEX Emission Line eXplorer (ELiXer) (Davis et al., in preparation), which uses multiple techniques including a Bayesian Lyman-$\alpha$-vs-[\ion{O}{2}] $\lambda\, 3727$ discrimination \citep{leung/etal:2017,farrow/etal:2021}.
To minimize contamination of nearby objects and artifacts, the ELiXer probability $P(\mathrm{Ly}\alpha)$ of every line in our sample must be larger than $0.9$ (the minimum $P(\mathrm{Ly}\alpha)/P($[\ion{O}{2}]$)$ is 9).

We require that the lines have $S/N\ge 6.5$ to minimize false positive detections.
The minimum throughput at $4540\,\angstrom$ of an observation must be $>0.08$ for HETDEX to include its line detections in the catalog.
In this work, we only include LAEs in observations with throughput $> 0.13$ and good seeing (PSF full-width-at-half-maximum (FWHM) $<1.4''$) to resolve the \Lya halos.
Furthermore, we only use observations that were taken in 2019 and later, as earlier data had more detector artifacts and often larger sky emission residuals.

After visually inspecting the remaining sources and excluding non-LAEs, our sample consists of 1491 high-confidence high-redshift \Lya emitting objects, which are detected in 150 observations.

\begin{figure*}
    \centering
    \includegraphics[width=0.9\textwidth]{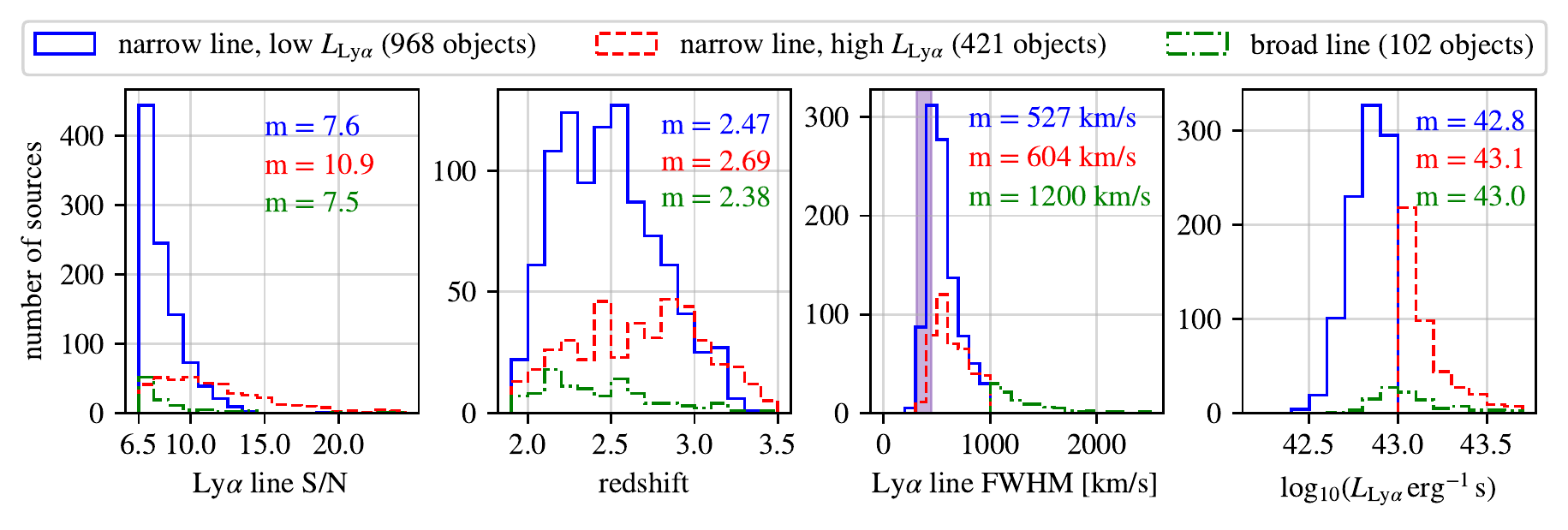}
    \caption{Distributions and median values (m) of several properties of the sources in the three samples: NLLL sources are shown in solid blue, NLHL sources are shown in dashed red, and BL sources are shown in dot-dashed green. The first panel presents the $S/N$ distributions for $S/N\ge 6.5$.
    the second panel displays the redshift distributions. The third panel shows the
    Lyman-$\alpha$-line FWHM. The purple shaded area shows the instrumental resolution over the redshift range of the instrument \citep[$5.6 \angstrom$;][]{hill/etal:2021}. The \textit{narrow-line} and \textit{broad-line} samples are separated at $\mathrm{FWHM}_{\mathrm{Ly}\alpha} = 1000\,\mathrm{km}\,\mathrm{s}^{-1}$. The fourth panel shows the \Lya luminosity. Because of the high-$S/N$ requirement, the sources have luminosities $>10^{42.4}\,\mathrm{erg}\,\mathrm{s}^{-1}$. The \textit{low-} and \textit{high-luminosity} samples are divided at $L_{\mathrm{Ly}\alpha} = 10^{43}\,\mathrm{erg}\,\mathrm{s}^{-1}$.}
    \label{fig:lae_sample_statistics}
\end{figure*}

This initial sample contains AGN\null. Since we are interested in the \Lya halos of LAEs without an AGN, we divide the sample into three subgroups. The first criterion is the line width: galaxies with a \Lya line $\mathrm{FWHM}_{\mathrm{Ly}\alpha}\geq 1000\,\mathrm{km}\,\mathrm{s}^{-1}$ are placed into the \textit{broad-line} sample (BL, 102 objects). The remaining sources are separated by their luminosity: galaxies with a \Lya luminosity $L_{\mathrm{Ly}\alpha}\geq 10^{43}\,\mathrm{erg}\,\mathrm{s}^{-1}$ constitute the \textit{narrow-line, high-luminosity} sample (NLHL, 421 objects), whereas galaxies with $L_{\mathrm{Ly}\alpha}< 10^{43}\,\mathrm{erg}\,\mathrm{s}^{-1}$ constitute the \textit{narrow-line, low-luminosity} (NLLL) sample (968 objects). We use the NLLL subset as our final LAE sample.
The luminosity threshold represents the luminosity at which narrow-band selected LAEs start to be dominated by AGN \citep{spinoso/etal:2020}. \citet{zhang/etal:2021} find that the AGN fraction at $2.0<z<3.5$ at $L_{\mathrm{Ly}\alpha}<10^{43}\,\mathrm{erg}\,\mathrm{s}^{-1}$ is $<0.05$; hence this is a conservative threshold.
In this fashion we remove most AGN from the NLLL/LAE sample.
We compare the median radial profile of the LAE sample to those of the other two samples to study the impact of potential AGN contamination.

\begin{figure}
    \centering
    \includegraphics[width=0.47\textwidth]{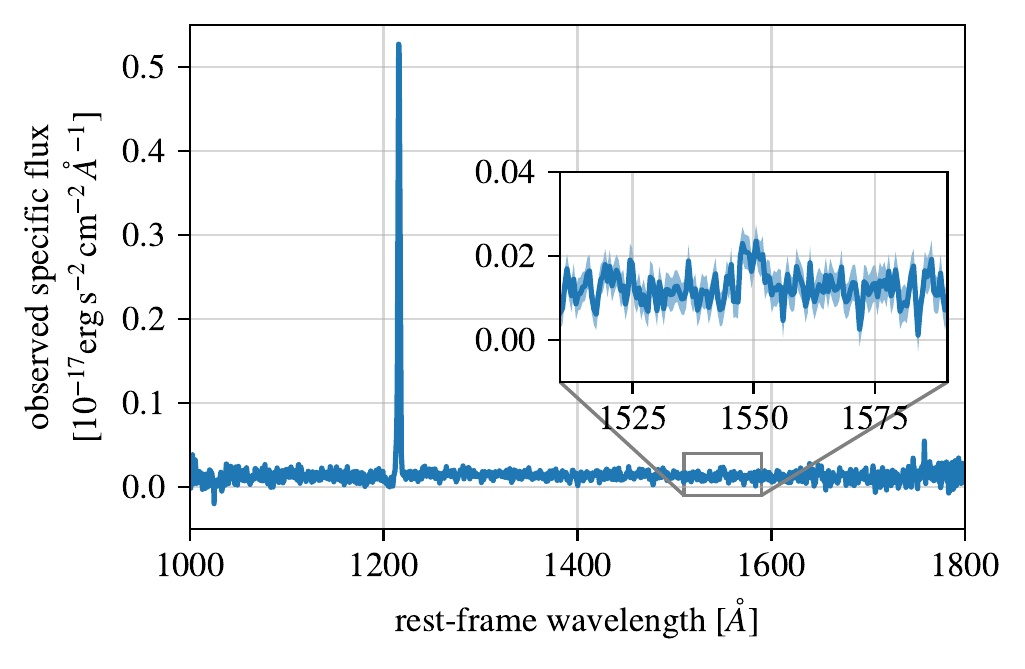}
    \caption{Median spectrum of the 968 LAEs in the NLLL sample in the rest frame. We use the original, not continuum-subtracted, spectrum of the fiber closest to each LAE. The observed specific flux is interpolated on a regular rest-frame wavelength grid with $0.4\,\angstrom$ binning to obtain the median spectrum. The inset shows the \ion{C}{4} emission line ($\mathrm{EW}_{\mathrm{C}\textsc{iv}}=4.0\pm0.7\,\angstrom$).}
    \label{fig:average_LAE_spectrum_restframe}
\end{figure}

Figure \ref{fig:lae_sample_statistics} shows the distributions of $S/N$, redshift, Lyman-$\alpha$-line FWHM, and  \Lya luminosities of the NLLL (blue), the NLHL (red), and the BL (green) samples.
Figure \ref{fig:average_LAE_spectrum_restframe} displays the median spectrum of the NLLL/LAEs in the rest frame without continuum subtraction. We interpolate the spectra of the closest fiber to each LAE on a rest-frame wavelength grid with a bin size of $0.4\,\angstrom$\null.
Then we take the median of the individual LAE spectra.

The median spectrum possesses continuum, a prominent \Lya line ($1216\,\angstrom$), and a faint \ion{C}{4} line ($1550\,\angstrom$). The large observed \Lya equivalent width ($\mathrm{EW}_{\mathrm{Ly}\alpha}=96\pm3\,\angstrom$) indicates that the LAE sample mainly consists of \Lya emitting galaxies, rather than low-$z$ [\ion{O}{2}] galaxies \citep{ciardullo/etal:2013,santos/etal:2020}. The strength of \ion{C}{4} emission ($\mathrm{EW}_{\mathrm{C}\textsc{iv}}=4.0\pm0.7\,\angstrom$) is consistent with predictions for star-forming galaxies \citep[][]{nakajima/etal:2018}. It is also within the range reported by \citet{feltre/etal:2020}, who studied the mean rest-frame UV spectra of LAEs at $2.9<z<4.6$ using MUSE\null.
Our NLHL sample has a similar \Lya equivalent width ($\mathrm{EW}_{\mathrm{Ly}\alpha}=94\pm3\,\angstrom$) and \ion{C}{4} emission ($\mathrm{EW}_{\mathrm{C}\textsc{iv}}=3.4\pm0.7\,\angstrom$). Our BL sample has a similar \Lya equivalent width ($\mathrm{EW}_{\mathrm{Ly}\alpha}=103\pm7\,\angstrom$), but larger \ion{C}{4} emission ($\mathrm{EW}_{\mathrm{C}\textsc{iv}}=5.5\pm1.5\,\angstrom$), as expected for a larger fraction of AGN in the sample.
The \ion{N}{5} ($1239\,\angstrom$ and $1243\,\angstrom$) and \ion{Si}{4} ($1394\,\angstrom$ and $1403\,\angstrom$) emission lines are not detected in median spectra of the NLLL, NLHL, or BL samples. The median spectra of the NLLL and NLHL samples have a \ion{He}{2} ($1640\,\angstrom$) equivalent width of $2.2\pm0.6\,\angstrom$, which is consistent with that of bright ($L_{\mathrm{Ly}\alpha}>10^{42.05}\,\mathrm{erg}\,\mathrm{s}^{-1}$) LAEs reported by \citet{feltre/etal:2020}.

\subsection{Masking and Continuum Subtraction}\label{subsec:masking_contsub}
We mask the wavelength regions around bright sky lines to avoid the largest sky emission residuals.

To isolate the \Lya emission, we remove continuum emission of the LAEs from the spectra. In each fiber, we subtract the median flux within $40\,\angstrom$ of the \Lya line center, but excluding the central $5\,\sigma_{\mathrm{Ly}\alpha}$, where $\sigma_{\mathrm{Ly}\alpha}$ is a Gaussian $\sigma$ from the fit to the emission line. Subtracting only the continuum on the red side of the \Lya line or changing the excluded central window from $5\sigma_{\mathrm{Ly}\alpha}$ to $7\sigma_{\mathrm{Ly}\alpha}$ does not affect our results.
We subtract the continuum emission of all fibers instead of masking those with high continuum emission for multiple reasons. Most importantly, the resulting surface brightness profiles have insignificant differences because the continuum subtraction successfully removes the continuum flux from projected neighbors. It is also difficult to mask continuum sources completely because of the large PSF of VIRUS. In the masking scheme, many fibers in the core of our LAE sample were masked. Finally, the continuum subtraction removes the systematic effect from an incorrect background subtraction, which can add a constant or smooth wavelength-dependent flux to each spectrum.

\section{Detection of Lyman-\texorpdfstring{$\alpha$ }\  halos}\label{sec:methods}

\subsection{Extraction of Lyman-\texorpdfstring{$\alpha$ }\ Surface Brightness}
\label{subsec:extraction_of_lya_sb}
We integrate the flux density over the wavelengths around the \Lya line of each LAE to obtain a surface brightness for each fibers that is located within $320\,\mathrm{kpc}$ of the LAE (typically 1100 fibers). The width of this integration window is different for each LAE and is chosen to be three times the $\sigma_{\mathrm{Ly}\alpha}$ of the Gaussian fit to the LAE's emission line. The integration window widths range from $5\,\angstrom$ to $18\,\angstrom$ (NLLL sample) in the observed reference frame, with a median (mean) of $10\,\angstrom$ ($12\,\angstrom$). To investigate the influence of the variable width on the radial profile measurement, the measurement was repeated with a fixed width of $\Delta\lambda=11\,\angstrom$; the results are consistent with one another. We choose the variable-width approach because the results have slightly higher S/N.

The result of this preparation is a set of surface brightness values as a function of angular separation from the centroid of each LAE. We translate this angular distance to a physical distance assuming our fiducial cosmology and sort the fibers around each LAE by their distance from it.

\subsection{Stacking}
We take the median surface brightness of all fibers around all LAEs in radial bins. The bin edges are at $[5, 10, 15, 20, 25, 30, 40,60,80,160,320]\,\mathrm{kpc}$. In each bin we gather all fibers whose distance from the fiber center to their corresponding LAE lies within this range. We do not normalize the surface brightness of the fibers around each LAE in order to retain physical units. We take the median surface brightness of these fibers and use a bootstrap algorithm to determine the uncertainty.
Each bin contains fibers from at least $55\%$ of LAEs in the NLLL sample, and at $r>10\,\mathrm{kpc}$, more than $96\%$ of the LAEs contribute to each bin.

\subsection{Estimating Systematic Uncertainty}\label{subsec:systematic_uncertainty}
We explore the systematic uncertainty in two steps.
The first part addresses the median surface brightness that would be interpreted as extended \Lya emission in random locations on the sky rather than centered on an LAE, which we refer to as the \textit{background} surface brightness. One total value is used for this quantity, which is the sum of contributions from \Lya emission in the target redshift, other redshifted emission lines, continuum emission from stars and nearby galaxies, and sky emission residuals.
To measure this background surface brightness, we repeat our surface brightness measurements at random locations. Specifically, for each LAE we randomly draw a fiber within the same observation and choose a random location within $1''$ of this fiber as the centroid.  To avoid the possibility of any of the LAE's halo \Lya affecting our experiment, we require that this new position be further than $2'$ from the original LAE\null. These random locations may coincide with foreground objects. This is intentional because we probe the \Lya halos out to $>30''$, which can include foreground objects. It is therefore appropriate not to exclude these from the random sample.
We then integrate the flux of the fibers within $20''$ of this location over the same wavelengths used for the integration of the
real LAE\null. For each LAE, we generate three of these random measurements, producing a data set for comparison that has the same distribution of wavelengths and widths of integration windows as the LAE, but is centered on random, but not necessarily empty, sky positions.
Because we prepare the spectra identically to those around LAEs, this background estimate accounts for potential systematic effects introduced by the continuum subtraction.

We use the median of our random position measurements to estimate the background surface brightness and a bootstrap algorithm to estimate the uncertainty.
We find $(4.0\pm 0.4)\times 10^{-20}\,\mathrm{erg}\,\mathrm{s}^{-1}\,\mathrm{cm}^{-2}\,\mathrm{arcsec}^{-2}$ for the NLLL sample.
The non-zero background does not contradict the efficacy of the full-frame sky subtraction. It can be caused for example by the complicated, asymmetric shape of the pixel flux distribution, non-flat fiber spectra, and differences in averaging estimators.

Having considered systematic uncertainties derived on larger volumes, we now address systematic errors associated with proximity of the LAEs.  This second step largely follows the procedure of \citet{wisotzki/etal:2018}. For each LAE we repeat the surface brightness extraction, but shift the central wavelength in increments of $10\,\angstrom$ from the observed \Lya wavelength, where the minimum offset is $20\,\angstrom$ and the maximum offset is $210\,\angstrom$. This produces 40 sets of Lyman-$\alpha$-free pseudo-narrow-band images for each LAE, which we then combine to make 40 Lyman-$\alpha$-free stacks, each separated by $\Delta\lambda$. Their standard deviation is then defined as the empirical uncertainty of the stack of LAEs. The median ratio of these empirical errors to the statistical error using the bootstrap algorithm is $1.5$. We adopt the larger of the two error estimates as the uncertainty of the median surface brightness in each bin.
The median of the wavelength-shifted profiles is consistent with the background at $r>50\,\mathrm{kpc}$.

We subtract the background surface brightness to find the median \Lya surface brightness profile around our LAE sample. Since the random sample depends on the sample of sources, the BL, NLHL, and NLLL samples each have one background surface brightness. The uncertainty of the background estimate is included in the uncertainty of the final \Lya surface brightness profile via Gaussian error propagation.

We can compare the reported errors to the propagation of errors from an individual fiber, which we expect to be smaller. The average flux uncertainty on an individual fiber is $\sim 7\times 10^{-18}\,\mathrm{erg}\,\mathrm{s}^{-1}\,\mathrm{cm}^{-2}$ within $10\,\angstrom$, which is the average integration width of the LAEs. Taking into account the fiber area and the number of fibers going into each bin, we can estimate the surface brightness limit for each bin. For example, at $120\,\mathrm{kpc}$, we have $3\times 10^5$ fibers, each with $1.5''$ diameter, which gives a surface brightness uncertainty of $7\times 10^{-21} \,\mathrm{erg}\,\mathrm{s}^{-1}\,\mathrm{cm}^{-2}\,\mathrm{arcsec}^{-2}$. Our measured uncertainty is $1.3\times10^{-20}\,\mathrm{erg}\,\mathrm{s}^{-1}\,\mathrm{cm}^{-2}\,\mathrm{arcsec}^{-2}$. The larger uncertainties are due to a combination of the intrinsic differences of LAEs and our attempt to include systematic effects.

\subsection{Shape of the Point Spread Function}

Multiple independent techniques show that the PSF of VIRUS is well modeled by a Moffat function with $\beta\in[3,3.5]$ in good seeing conditions \citep{hill/etal:2021,gebhardt/etal:2021}. To test this, we measure the median radial profile of \textit{Gaia} stars observed by VIRUS in the same observations as the LAEs.
We select 3795 faint stars (\textit{g}-band magnitude between 19 and 20) from the \textit{Gaia} DR1 catalog \citep{gaia/etal:2016}.
For each fiber close to the centroid of the star, we compute the weighted mean flux density within $4550\,\angstrom<\lambda<4650\,\angstrom$, where the weights are the inverse squared flux density errors. We use the spectra without continuum subtraction for the stars, since their spectra mostly consist of continuum emission.
The radial profiles of the stars are stacked by taking the median in radial bins.

\begin{figure}
    \centering
    \includegraphics[width=0.47\textwidth]{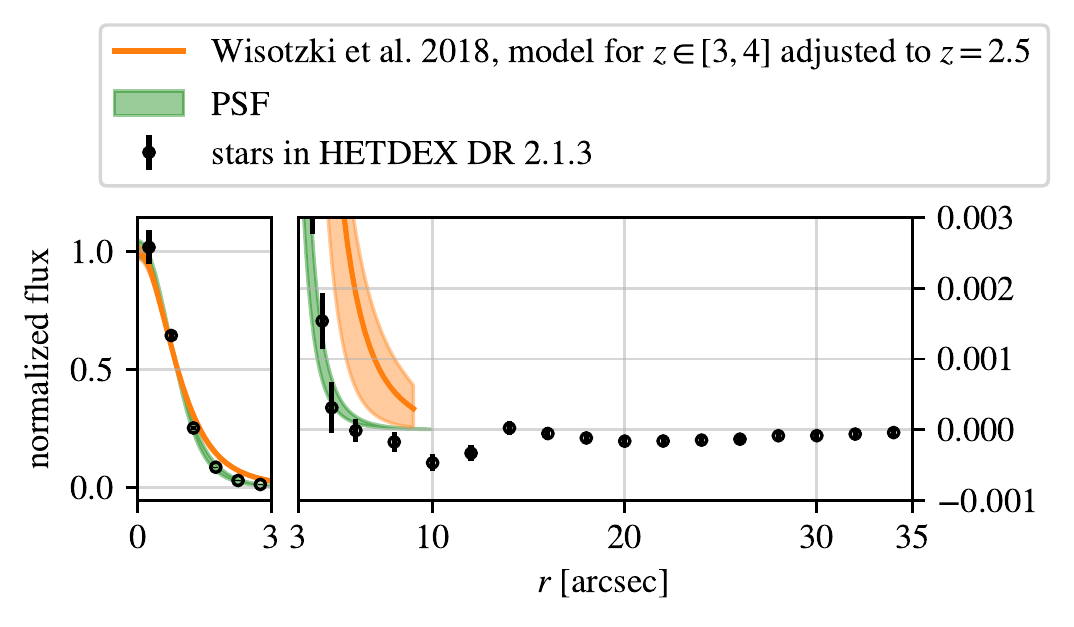}
    \caption{Normalized median radial profile of 3795 stars in our observations (black points) compared to the PSF model (green). The model is a Moffat function with $\beta=3$ and $1.2''\leq \mathrm{FWHM} \leq 1.4''$. The orange shaded area is the best-fit model from \citet{wisotzki/etal:2018} for \Lya halos at $3<z<4$ adjusted to our observation and normalized.
    The right panel is a continuation of the left panel in radius but covering a smaller range in flux.  The radial profile of the stars is negative at $r\gtrsim 5''$ because of small errors in the background light correction.}
    \label{fig:star_stack_vs_psf}
\end{figure}

Figure \ref{fig:star_stack_vs_psf} shows the stacked radial profile of the stars compared to the PSF model with $\beta=3$. The exact choice of $\beta$ does not affect our results. While the profile traces the PSF shape well at $r\lesssim 5''$, the median flux at larger radii is mostly negative.
We suspect that this behavior is due to an over-estimating of the sky from undetected background galaxies. We correct this effect for the LAEs by using the continuum-subtracted spectra and by subtracting the background surface brightness.

\section{Results}\label{sec:results}

\begin{figure*}
    \centering
    \includegraphics[width=0.47\textwidth]{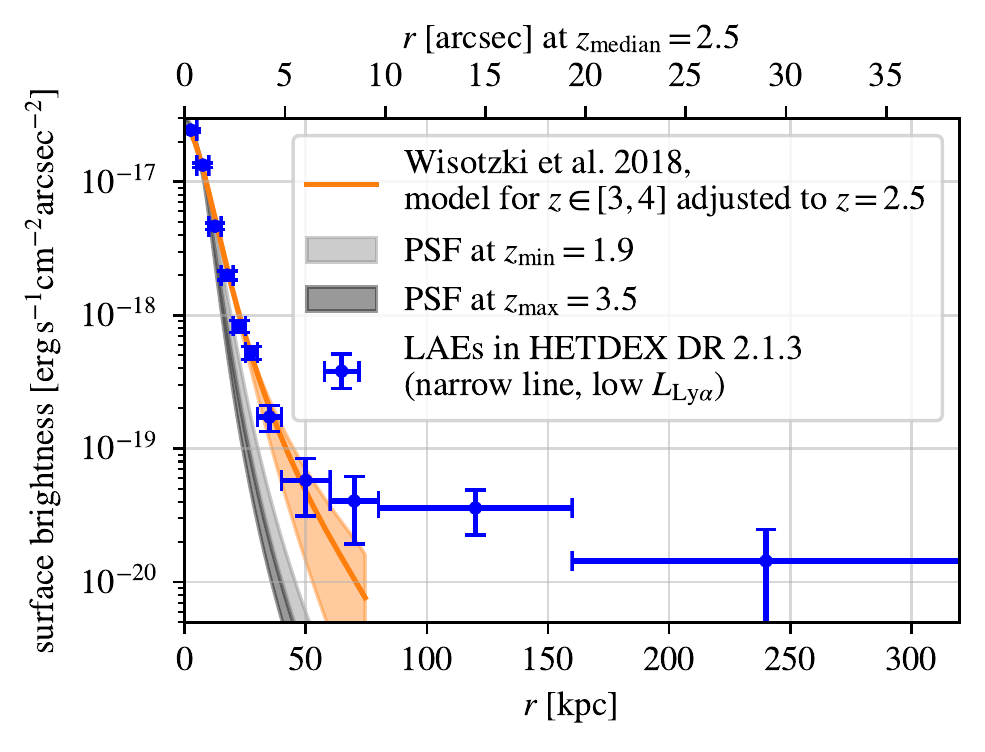}
    \includegraphics[width=0.47\textwidth]{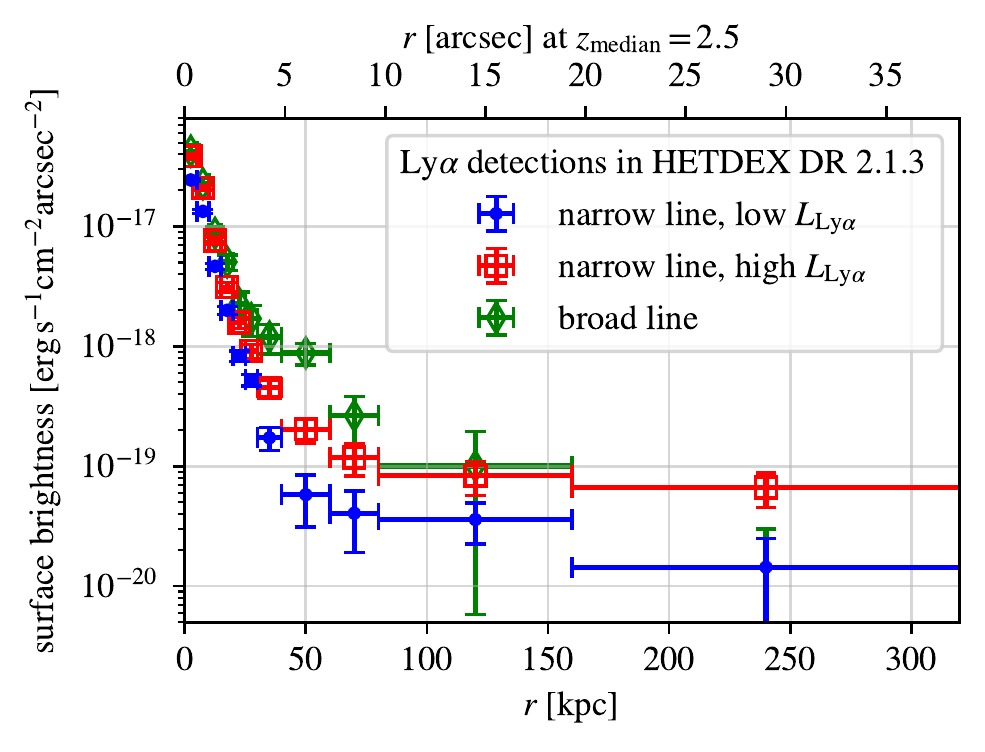}
    \caption{\textbf{Left:} Median \Lya surface brightness profile of our LAE sample (blue, $\mathrm{FWHM}_{\mathrm{Ly}\alpha}<1000\,\mathrm{km}\,\mathrm{s}^{-1}$ and $L_{\mathrm{Ly}\alpha}<10^{43}\,\mathrm{erg}\,\mathrm{s}^{-1}$) after subtracting the background surface brightness. The gray shaded area shows the PSF in the observations at the minimum and maximum redshift. The orange profile is the best-fit model in \citet{wisotzki/etal:2018} for LAEs at
    $3<z<4$, adjusted to our observations and rescaled to match the flux of our galaxies in the core. Our measured radial profile agrees well with this model within $80\,\mathrm{kpc}$, but is flatter at larger radii.
    \textbf{Right:} Comparison of the median \Lya surface brightness profile of the LAE sample (blue) to those with broad lines (green diamonds) and narrow lines and high luminosities (red squares). The background surface brightness values of each sample have been subtracted.}
    \label{fig:rad_prof_contsub}
\end{figure*}

The left panel of Figure \ref{fig:rad_prof_contsub} displays the median \Lya surface brightness profile around our set of LAEs (blue, NLLL sample). The PSF in our observations is shown as a comparison. This median \Lya surface brightness profile is clearly more extended than the profile of a point source and shows significant emission of $(3.6\pm 1.3)\times 10^{-20}\,\mathrm{erg}\,\mathrm{s}^{-1}\,\mathrm{cm}^{-2}\,\mathrm{arcsec}^{-2}$ out to $160\,\mathrm{kpc}$.
The median profile of continuum emission at longer wavelengths than \Lya is consistent with the PSF, but it has insufficient S/N for a meaningful comparison.

The right panel of Figure \ref{fig:rad_prof_contsub} shows the comparison of the median radial profiles of the three samples: the LAE/NLLL sample (blue), the NLHL sample (red) and the BL sample (green). The \textit{narrow-line} samples have similar shapes, while the BL sample is flatter at intermediate radii ($25\,\mathrm{kpc}\leq r\leq 60\,\mathrm{kpc}$). Both the NLHL and the BL samples have a higher overall surface brightness than the LAE/NLLL sample, suggesting that the effect of potential AGN contamination in the LAE sample is a higher overall surface brightness and possibly a flattening of the radial profile at intermediate radii. For example, the stacked surface brightness profile of 15 quasars at $z\sim 2$ obtained from narrow-band images yields $(5.5\pm3.1)\times10^{-20}$~erg s$^{-1}$ cm$^{-2}$ arcsec$^{-2}$ in a large radial bin of $50 < r < 500$~kpc \citep{arrigonibattaia/etal:2016}. While the statistical significance is low, this level of surface brightness is consistent with that of our NLHL sample. 

The left panel of Figure \ref{fig:rad_prof_contsub} compares the profile to the best-fit model for the stack of LAEs at $3<z<4$ reported in \citet{wisotzki/etal:2018}. This model consists of a point-like profile proportional to the PSF plus a halo profile following a S\'{e}rsic function. The point-like contribution has a total flux $F_\mathrm{ps} = \left(232\pm50\right)\times 10^{-20}\,\mathrm{erg}\,\mathrm{s}^{-1}\,\mathrm{cm}^{-2}$ and $\mathrm{FWHM}=0.703''$ for the MUSE instrument. The S\'{e}rsic function has the total flux $F_\mathrm{h}=\left(1488\pm83\right)\times 10^{-20}\,\mathrm{erg}\,\mathrm{s}^{-1}\,\mathrm{cm}^{-2}$, effective radius $r_\mathrm{eff,h}=0.86''\pm 0.11''$, and S\'{e}rsic index $n_\mathrm{h} = 2.8\pm1.1$.

Before we compare this model with our data, we have to account for several differences between the two data sets. First, our LAE sample is located at lower redshifts ($1.9< z < 3.5$) and it is on average much brighter (by a factor of $10$ in \Lya luminosity) in \Lya than the LAE sample of \citet{wisotzki/etal:2018}. Second, VIRUS has a larger PSF and larger fiber diameter than MUSE. To account for the redshift difference we translate the angular separation at $z_\mathrm{mid}^\mathrm{MUSE}=3.5$ to physical distances and back to the corresponding angular separation at our median redshift $z_\mathrm{median}=2.5$ ($10\%$ change in scale). We then convolve the model profile with the median PSF of our observations, i.e., a Moffat function with $\beta=3$ and $\mathrm{FWHM}=1.3''$. Since the fibers on VIRUS are larger, we also convolve the model profile with the fiber face (a tophat with radius $0.75''$). As our sample of galaxies is brighter, we multiply the model profile by $10.3$ such that the flux in the core ($r\leq 2''$) of our measured radial profile and the model match. 

The adjusted model agrees qualitatively well with our measured radial profile at $r<80\,\mathrm{kpc}$ despite the differences in redshift and \Lya luminosities of the LAE samples. At larger radii, our radial profile becomes flatter than the model. However, the measured radial profile from \citet{wisotzki/etal:2018} is also above the fitted profile by $2\,\sigma$ at $60\,\mathrm{kpc}$, which is consistent with a flatter outer radial profile.

There are three possible reasons for the discrepancy at larger radii. First, an unknown systematic error in our analysis may artificially flatten the radial profile. Second, the smaller field of view of MUSE may cause them to over-subtract extended \Lya emission and thereby miss the flattening. Third, the flattening may depend on the \Lya luminosity or redshift of the LAE sample and may be stronger for brighter, probably more massive LAEs at lower redshift.
To test the second possible reason, we imitate a background subtraction for each VIRUS IFU, which is slightly smaller than the MUSE field of view. This removes the flattening at $r\geq 50\,\mathrm{kpc}$.
To investigate a possible luminosity dependence, future studies can expand this analysis to fainter LAEs in HETDEX data. The redshift dependence can be tested by expanding the LAE sample and splitting it into redshift bins.
\section{Discussion}
\label{sec:discussion}

\subsection{Impact of the Background Subtraction}\label{subsec:background_subtractions}
The main goal of this paper is to measure the shape of the \Lya emission around LAEs, which requires the removal of background emission.
We subtract background emission in three steps: the sky subtraction (Section \ref{subsec:data_processing}), the continuum subtraction (Section \ref{subsec:masking_contsub}), and the subtraction of the remaining background surface brightness (Section \ref{subsec:systematic_uncertainty}).
The sky subtraction removes the (biweight) average flux within $18'$, which covers $8-9\,\mathrm{Mpc}$ in our redshift range, in each $2\,\angstrom$ wavelength bin.
The continuum subtraction removes the median flux within $\Delta\lambda=80\,\angstrom$ in the observed frame, which corresponds to a line-of-sight distance of $12-36\,\mathrm{Mpc}$, of each fiber.
The background subtraction removes the median surface brightness of random locations within $18'$ or $8-9\,\mathrm{Mpc}$ with the same wavelength and integration-window distribution as the LAEs.

It is difficult to disentangle contributions from sky emission residuals, astronomical foreground objects, and the genuine diffuse \Lya background to the subtracted background estimates.
Due to these background subtraction procedures we may therefore underestimate the \Lya surface brightness.
The affected scale is $\sim 10\,\mathrm{Mpc}$, which is almost two orders of magnitude larger than our observed range of the \Lya profiles. Hence the uncertainty of the background \Lya surface brightness should manifest itself as a constant additive term and does not affect the shape of the \Lya profiles.

\subsection{Comparison with Theoretical Predictions}
\begin{figure}
    \centering
    \includegraphics[width=0.47\textwidth]{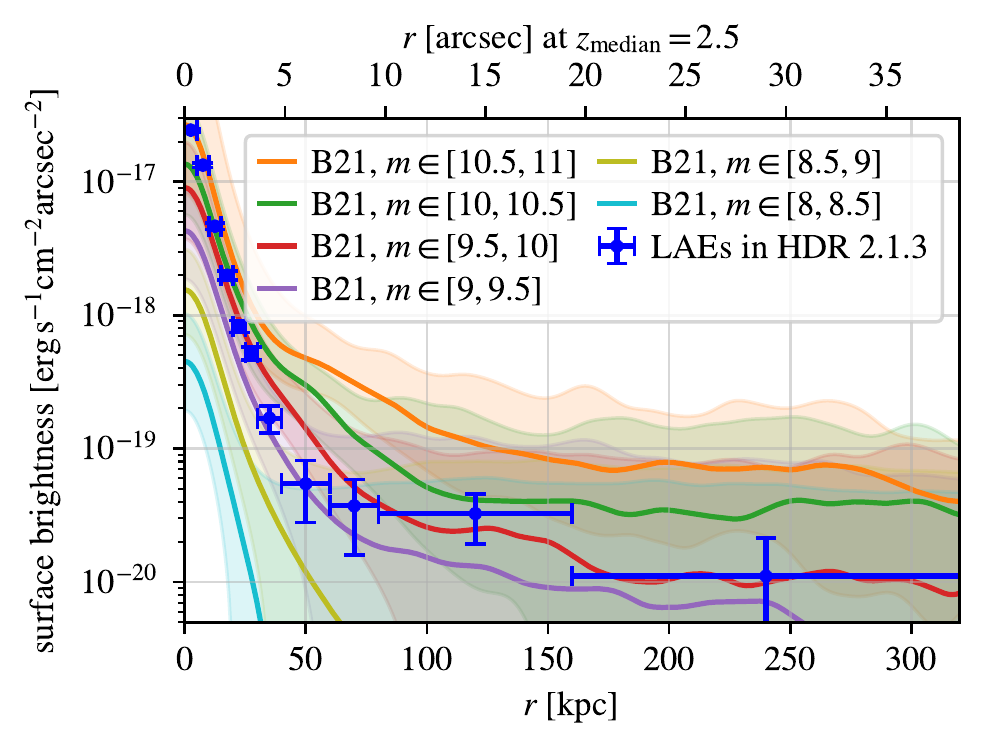}
    \caption{Median \Lya surface brightness of the NLLL LAE sample (blue) after subtracting the background, compared to the median simulated surface brightness profiles in \citet{byrohl/etal:2021} in six stellar mass bins ($m=\log_{10}(M_\star/M_\odot)$), adjusted to our observations.}
    \label{fig:chris_comparison_plot_contsub}
\end{figure}

Figure \ref{fig:chris_comparison_plot_contsub} compares the median radial profile of the NLLL LAE sample with the prediction for \Lya halos from the simulation in \citet{byrohl/etal:2021} at $z=3$. 
The surface brightness of the simulated \Lya halos is integrated over $2.2\,\mathrm{Mpc}$-wide slices along the line of sight. This approach is similar to the integration width of our measurement, which has a median (mean) of $2.7\,\mathrm{Mpc}$ ($2.9\,\mathrm{Mpc}$).
To adjust the radial profile from the simulation to our observations, we convolve it with the median PSF and VIRUS fiber face. We then subtract the mean \Lya surface brightness in the entire simulation volume ($1.9\times 10^{-20}\,\mathrm{erg}\,\mathrm{s}^{-1}\,\mathrm{cm}^{-2}\,\mathrm{arcsec}^{-2}$) from the simulated radial profiles to emulate the background subtraction in our data analysis. Finally, we multiply the profiles by $(1+3)^4/(1+2.5)^4\approx 1.7$ to account for surface brightness dimming.

\begin{figure*}
    \centering
    \includegraphics[width=0.47\textwidth]{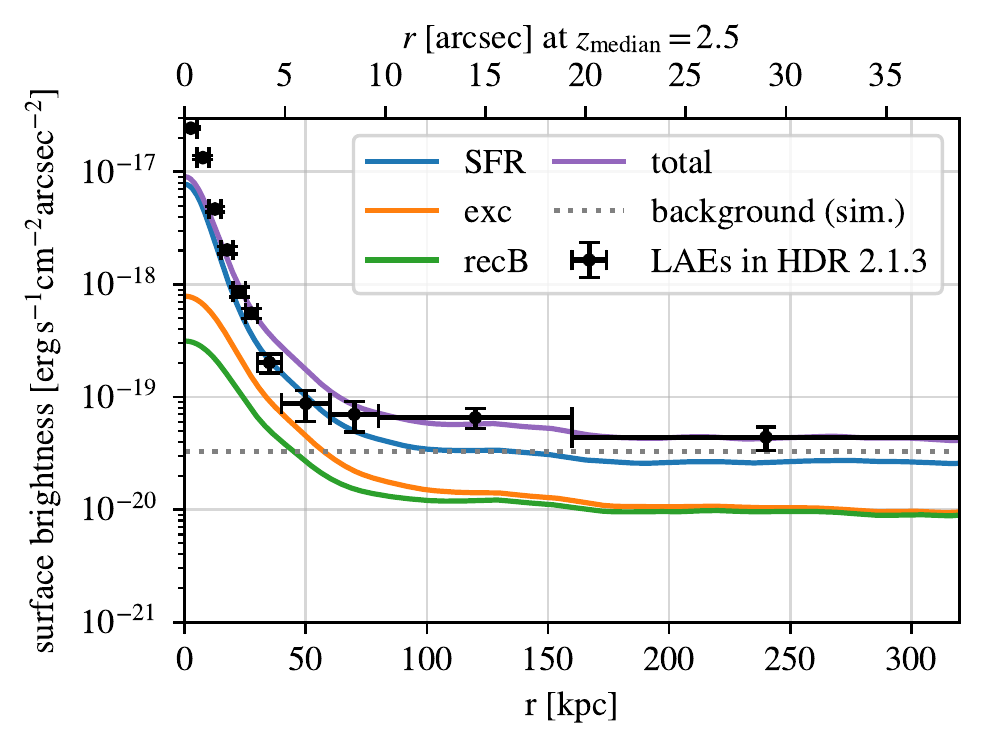}
    \includegraphics[width=0.47\textwidth]{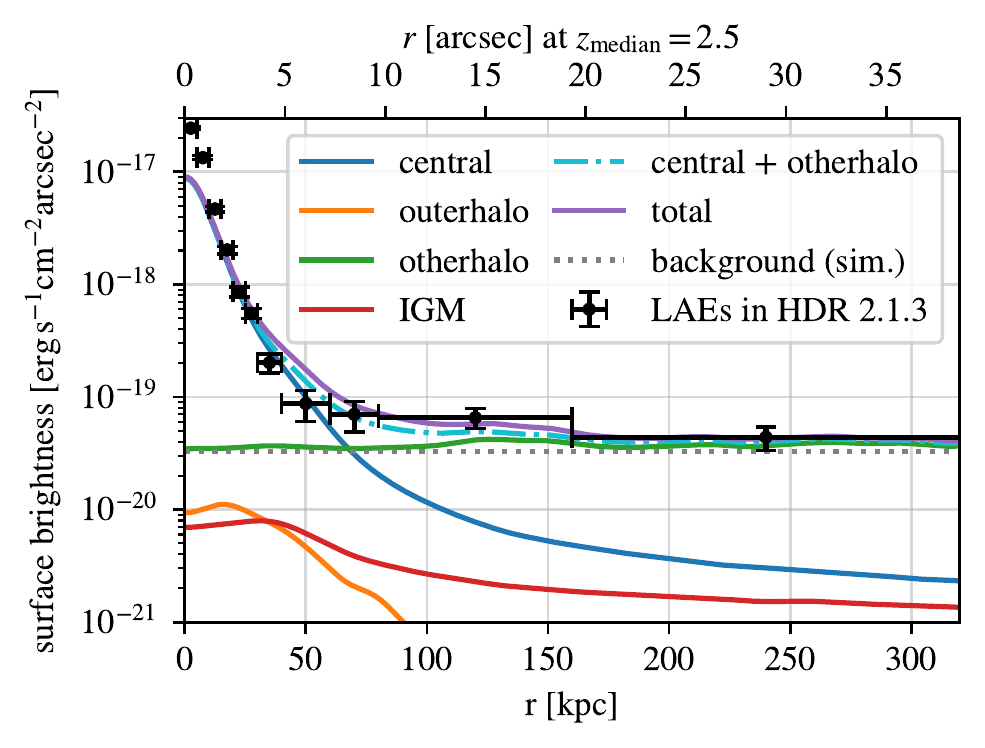}
    \caption{
    Median \Lya surface brightness profile of the NLLL/LAE sample in HETDEX (black) compared to the median surface brightness profiles of simulated galaxies with stellar masses $M_\star\in[10^{9.5}M_\odot, 10^{10}M_\odot]$ from \citet{byrohl/etal:2021}. The dashed gray line indicates the mean \Lya surface brightness of the simulation (background (sim.)), which we add to the measured \Lya surface brightness profile.
    \textbf{Left:} Different emission mechanisms of the \Lya photons: recombination after photoionization from star formation (SFR, blue), collisional excitation (exc, orange), case-B recombination after including photoionization from an ionizing background (recB, green), and the sum of these (total, purple). SFR dominates the simulated profile at all radii.
    \textbf{Right:} Different emission origins of the \Lya photons: the central galaxy in the target dark matter halo (central, blue), the outer parts of this dark matter halo (outerhalo, orange), another dark matter halo than the target halo (otherhalo, green), IGM (red), and the sum of these (total, purple). The sum of the central and otherhalo contributions (cyan dot-dashed line) reproduces the measured surface brightness profile well except in the core ($r<10\,\mathrm{kpc}$).
    }
    \label{fig:stack_vs_mechanisms_chris_105_110}
\end{figure*}
This figure displays the median radial profiles of simulated galaxies in six stellar mass ranges between $10^8 M_\odot$ and $10^{11} M_\odot$. Our measurements are consistent with the simulated ones; in detail, however, our measured \Lya surface brightness profile is steeper than the profiles from the simulation at $r\lesssim 50\,\mathrm{kpc}$.

The \citet{byrohl/etal:2021} simulation finds that most photons illuminating \Lya halos originate from star-forming regions within the central galaxy or, at large radii, nearby galaxies, which are scattered in the CGM/IGM. Other emission mechanisms and emission origins contribute less to the total median \Lya profile. Figure \ref{fig:stack_vs_mechanisms_chris_105_110} compares the measured surface brightness profile with the median profiles of individual emission mechanisms and emission origins from simulated galaxies with $M_\star\in[10^{9.5}M_\odot, 10^{10}M_\odot]$. For an informative comparison, the mean Lya surface brightness from the simulation (``background (sim.)'') is added to the measured profile, as the mean surface brightness for individual simulated components was not available.

The left panel shows different emission mechanisms. SFR dominates the simulated surface brightness profile at all radii.

The right panel presents different emission origins before scattering of the \Lya photons. The sum of the central component and the otherhalo component is close to the measured surface brightness profile, suggesting that photons originating from the outer parts of the dark matter halo and IGM play only a minor role in forming the total profile.

While the level of agreement is impressive, there are small discrepancies. One reason is linked to modeling limitations in the hydrodynamic simulations and their \Lya radiative transfer treatment: the lack of coupled ionizing radiation from star-forming regions, the lack of dust modeling, and uncertainty in the intrinsic \Lya luminosities within galaxies.

The \Lya radiation escaping from the ISM is largely determined by the complex radiative transfer in the ISM's multiphase state that remains unresolved in the simulation. Along with large uncertainties on the intrinsic \Lya luminosity of stellar populations and potential \Lya emission from obscured AGN, large uncertainties for the \Lya luminosity escaping the ISM exist. The assumed linear scaling between star-formation rate and ISM-escaping \Lya luminosity in~\cite{byrohl/etal:2021} may not accurately reflect reality. The observations reveal substantial scatter in the scaling relation 
\citep{santos/etal:2020,runnholm/etal:2020}. The selection of the galaxies in the observation is also different than in the simulation.
We may therefore not reflect a potential bias for the observational sample regarding their star-formation and dust content. Finally, while the simulation uses the mean to convert two-dimensional images to radial profiles, we use the median.

In addition to the comparison to hydrodynamical simulations, we also consider an analytically motivated approach by \citet{kakiichi/etal:2018}, who predict a \Lya radial profile following a power law $\propto r^{-2.4}$ for $r\in [20,1000]\,\mathrm{kpc}$. In this approach, only scatterings of photons from the central galaxy in the CGM are considered. We compare this prediction with our measurement by fitting a similar model to our stack. The model consists of a point-like component given by a $\delta$ function times a constant $a$, plus a power-law halo, which we terminate at $r_0=1''$ ($\approx 8\,\mathrm{kpc}$) to avoid divergence at smaller radii.
The profile is therefore given by
\begin{equation}
    f(r, \phi) = a \times \delta(r) + b \times \begin{cases}
    1 & \text{if } r<1''\\
    \left(\frac{r}{''}\right)^{-2.4} & \text{if } r\geq 1''.
\end{cases}
\end{equation}
We convolve the profile at the median redshift ($z_\mathrm{median}=2.5$) with the PSF and VIRUS fiber face and fit to the data by varying the constants $a$ and $b$. 

\begin{figure}
    \centering
    \includegraphics[width=0.47\textwidth]{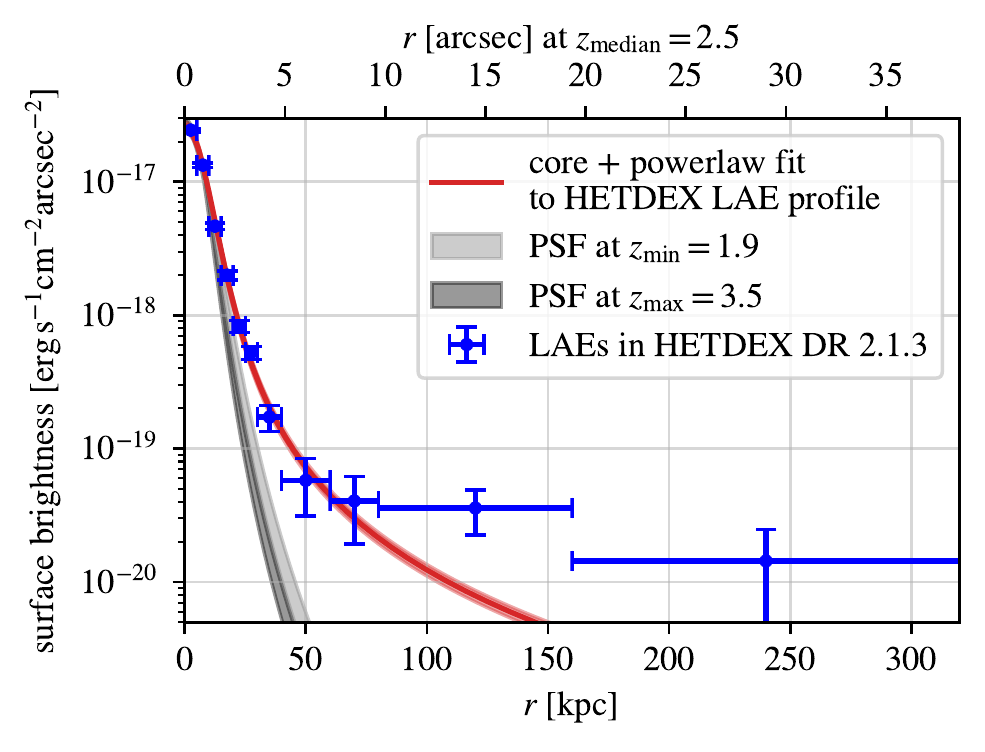}
    \caption{Median \Lya surface brightness of the NLLL LAE sample (blue) after subtracting the background. The gray shaded area represents the PSF in the observations at the minimum and maximum redshift. The red line is the least-squares fit of a ``core plus power-law'' profile motivated by \citet{kakiichi/etal:2018}. 
    The red shaded area shows the 1-$\sigma$ uncertainty of the least-squares fit.}
    \label{fig:powerlaw_fit_contsub_stack_compare}
\end{figure}

Figure \ref{fig:powerlaw_fit_contsub_stack_compare} shows the result. The power-law fit agrees well with the data out to $r = 80\,\mathrm{kpc}$. At larger radii, it underestimates the \Lya surface brightness.
Since the model only considers scattering of \Lya photons originating from the central galaxy, the flattening of the profile at large radii ($r\gtrsim 100\,\mathrm{kpc}$) may be caused by photons that originate from other dark matter halos, as predicted by \citet{byrohl/etal:2021}. The outer surface brightness profile is also consistent with the results of \citet{mas-ribas/dijkstra:2016} using the CGM model of \citet{dijkstra/kramer:2012}, further suggesting that the clustering of ionizing sources around the LAEs is an important factor.

Another reason for the flattening of the \Lya profile may be fluorescence from the ultraviolet background. The values predicted by \citet{cantalupo/etal:2005} and \citet{gallego/etal:2018,gallego/etal:2021} are on the order of $10^{-20}\,\mathrm{erg}\,\mathrm{s}^{-1}\,\mathrm{cm}^{-2}\,\mathrm{arcsec}^{-2}$ at $z\sim 3$, which is consistent with the simulations of~\cite{byrohl/etal:2021} and the outermost points of our observed profile, but insufficient to explain the surface brightness at intermediate radii. Recent observations hint at the detection of the filamentary structure of the cosmic web traced by \Lya at a surface brightness $\sim 10^{-20}-10^{-18}\,\mathrm{erg}\,\mathrm{s}^{-1}\,\mathrm{cm}^{-2}\,\mathrm{arcsec}^{-2}$ at $z\sim 3$ in overdense regions \citep[][]{umehata/etal:2019,arrigonibattaia/etal:2019b,bacon/etal:2021}. Because of their rarity and small volume, they may contribute to, but are unlikely to dominate the median-stacked profile on large scales. Future multi-dimensional analysis similar to \citet[][]{leclercq/etal:2020} can give additional insights on the nature of the extended emission and its link to the Lyman-$\alpha$-emitting cosmic web.

Finally, satellite galaxies or galaxies within the integration window of the \Lya line along the line of sight may contribute to the extended \Lya emission \citep{mas-ribas/etal:2017}.

\section{Summary}\label{sec:summary}
We presented the median radial \Lya surface brightness profile of 968 LAEs at $1.9<z<3.5$ that were carefully selected from the DR 2.1.3 of the HETDEX survey. The presence of \Lya halos is detected at $(3.6\pm 1.3)\times 10^{-20}\,\mathrm{erg}\,\mathrm{s}^{-1}\,\mathrm{cm}^{-2}\,\mathrm{arcsec}^{-2}$ out to $160\,\mathrm{kpc}$. 
The potential residual AGN contamination in the LAE sample may increase the overall amplitude of the median radial profile and may flatten the profile in the intermediate radii.

We compared the radial profile with the rescaled model of \citet{wisotzki/etal:2018} for the median radial profile of fainter LAEs at $3<z<4$, which we adjusted to the VIRUS observations. This adjusted model agrees well with our radial profile at $r\lesssim 80\,\mathrm{kpc}$. At larger radii, our measured profile is flatter.

We also compared the radial profile with the median radial \Lya surface brightness profiles of galaxies with stellar masses of $10^{8}-10^{11}M_\odot$ at $z=3$, taken from the radiative transfer simulation of \citet{byrohl/etal:2021}. The simulation results agree well with our measurement at most radii, except at $r<10\,\mathrm{kpc}$ and $30\,\mathrm{kpc}<r<60\,\mathrm{kpc}$.
The comparison suggests 
that our surface brightness profile at $r\lesssim100$~kpc is dominated by photons emitted in star-forming regions in the central galaxy and, at $r\gtrsim 100\,\mathrm{kpc}$, by photons from galaxies in other dark matter halos. 
A similar conclusion was reached in \cite{kikuchihara/etal:2021}.

Finally, we compared the radial profile with the prediction of \citet{kakiichi/etal:2018} that the \Lya halo is proportional to $r^{-2.4}$ for $r>20\,\mathrm{kpc}$. This power-law profile fits the measured radial profile well at $r\leq 80\,\mathrm{kpc}$. At larger radii our measured profile is flatter. Since \citet{kakiichi/etal:2018} only considered scattered photons from the central galaxy, this result further suggests that the flattening is due to photons originating from other dark matter halos.

The radiative transfer simulation of \citet{byrohl/etal:2021} predicts that the median \Lya profiles of LAEs have similar shapes across redshifts, stellar masses, and luminosities. The similarity of the shapes of our median radial profile and that in \citet{wisotzki/etal:2018}, which are at larger redshifts and fainter than our sample, suggests that median \Lya profiles at small radii indeed have similar shapes between redshifts $2<z<4$ and across a factor of $10$ in luminosity. 

In conclusion, this measurement of faint \Lya surface brightness to $>100$~kpc from LAEs shows the high scientific potential of HETDEX observations. The methods to quantify systematic uncertainties developed in this paper will be valuable for \Lya intensity mapping \citep{kovetz/etal:2017,croft/etal:2018,kakuma/etal:2021,kikuchihara/etal:2021} with HETDEX, which will improve constraints on cosmological parameters.

\acknowledgments
We thank M. Gronke for useful discussions and F. Arrigoni Battaia for comments on the draft. We thank the reviewer for their helpful, constructive feedback.
EK's work was supported in part by the Deutsche Forschungsgemeinschaft (DFG, German Research Foundation) under Germany's Excellence Strategy - EXC-2094 - 390783311.

HETDEX is led by the University of Texas at Austin McDonald Observatory and Department of Astronomy with participation from the Ludwig-Maximilians-Universit\"at M\"unchen, Max-Planck-Institut f\"ur Extraterrestrische Physik (MPE), Leibniz-Institut f\"ur Astrophysik Potsdam (AIP), Texas A\&M University, Pennsylvania State University, Institut f\"ur Astrophysik G\"ottingen, The University of Oxford, Max-Planck-Institut f\"ur Astrophysik (MPA), The University of Tokyo, and Missouri University of Science and Technology. In addition to Institutional support, HETDEX is funded by the National Science Foundation (grant AST-0926815), the State of Texas, the US Air Force (AFRL FA9451-04-2-0355), and generous support from private individuals and foundations.

The observations were obtained with the Hobby-Eberly Telescope (HET), which is a joint project of the University of Texas at Austin, the Pennsylvania State University, Ludwig-Maximilians-Universität München, and Georg-August-Universität Göttingen. The HET is named in honor of its principal benefactors, William P. Hobby and Robert E. Eberly.

VIRUS is a joint project of the University of Texas at Austin, Leibniz-Institut f{\" u}r Astrophysik Potsdam (AIP), Texas A\&M University (TAMU), Max-Planck-Institut f{\" u}r Extraterrestrische Physik (MPE), Ludwig-Maximilians-Universit{\" a}t M{\" u}nchen, Pennsylvania State University, Institut f{\" u}r Astrophysik G{\" o}ttingen, University of Oxford, Max-Planck-Institut f{\" u}r Astrophysik (MPA), and The University of Tokyo.

The authors acknowledge the Texas Advanced Computing Center (TACC) at The University of Texas at Austin for providing high performance computing, visualization, and storage resources that have contributed to the research results reported within this paper. URL: http://www.tacc.utexas.edu.

The Kavli IPMU is supported by World Premier International Research Center Initiative (WPI), MEXT, Japan. The Institute for Gravitation and the Cosmos is supported by the Eberly College of Science and 
the Office of the Senior Vice President for Research at Pennsylvania State University.

This research made use of NASA's Astrophysics Data System Bibliographic Services.

This research has made use of the VizieR catalogue access tool, CDS, Strasbourg, France.  The original description of the VizieR service was published in A\&AS 143, 23.

This research made use of Astropy,\footnote{http://www.astropy.org} a community-developed core Python package for Astronomy \citep{astropy:2013, astropy:2018}, Matplotlib \citep{Hunter:2007}, Numpy \citep{harris2020array}, and Scipy \citep{2020SciPy-NMeth}.

%

\facilities{The Hobby-Eberly Telescope (McDonald Observatory)}

\bibliography{references}{}
\bibliographystyle{aasjournal}



\end{document}